\documentclass[aps,pra,twocolumn,superscriptaddress]{revtex4-1}

\usepackage{amssymb}
\usepackage{graphicx}
\usepackage{amsmath}

\usepackage{hyperref}

\begin{document}
	
	\title{Evaporation dynamics of the Sachdev-Ye-Kitaev model}
	\author{Pengfei Zhang}
	\affiliation{Institute for Quantum Information and Matter and Walter Burke Institute for Theoretical Physics, California Institute of Technology, Pasadena, California 91125, USA}
	\affiliation{Institute for Advanced Study, Tsinghua University, Beijing, 100084, China}
	\date{\today}
	
	\begin{abstract}
	In this paper, we study the evaporation dynamics of the Sachdev-Ye-Kitaev model, with an initial temperature $T_\chi$, by coupling it to a thermal bath with lower temperature $T_\psi<T_\chi$ modeled by a larger SYK model \cite{Yiming}. The coupling between the small system and the bath is turned on at time $t=0$. Then the system begins to envolve and finally becomes thermalized. Using the Keldysh approach, we analyze the relaxation process of the system for different temperatures and couplings. For marginal or irrelevant coupling, after a short-time energy absorption, we find a smooth thermalization of the small system where the energy relaxes before the system become thermalized. The relaxation rate of effective temperature is found to be bounded by $T$, while the energy thermalization rate increases without saturation when increasing the coupling strength. On the contrary, for the relevant coupling case, both energy and effective temperature show oscillations. We find this oscillations frequency to be coincident with the excitation energy of a Majorana operator.
	\end{abstract}
	
	\maketitle
	\section{Introduction}
In recent years, the non-equilibrium dynamics of quantum many-body systems has drawn a lot of attention. As interesting progress, inspired by gravitational calculations, it is understood that for isolated quantum systems, there is an upper bound for relaxation rate $\Gamma\leq O(1/\beta)$ \cite{Hqm,thermal}. For strongly interacting systems, this bound is approximately saturated, and the relaxation rate corresponds to the typical decay rate of quasi-normal modes in the gravity description for holographic models \cite{gra}. 

While most of these studies for the quantum dynamics of strongly interacting systems focus on isolated systems, the relaxation of quantum systems coupled to a bath should also be an interesting problem. On the one hand, in real materials, the system is inevitably open due to the coupling to phonons \cite{kittel,simons}. On the other hand, the coupling to quantum fields would give rise to interesting physics. As an example, coupling a black hole to quantum fields give would rise to the celebrated Hawking radiation \cite{Hawking,carroll}. A further coupling to the thermal bath shows a possible resolution of the black-hole information paradox \cite{para1,para2,para3}. Motivated by these results, in this paper, we would like to study such quantum dynamics of strongly correlated many-body systems coupled to an external quantum bath. 

Generally, the real-time evolution of quantum systems can be formulated in terms of a path integral on Keldysh contour \cite{Kamenev,book}, where the two-point functions are determined by Kadanoff-Baym equations \cite{book}. However, for a strongly interacting system with possible holographic interpretation, these set of equations are usually hard to be solved with controlled accuracy, due to the lack of small parameters. 

Fortunately, the Sachdev-Ye-Kitaev (SYK) model proposed by Kitaev \cite{Kitaev2} in recent years based on early works by Sachdev and Ye \cite{Ye}, turns out to be an ideal platform for the study of both field theoretical \cite{Kitaev2,Comments,spectrum1,spectrum2,spectrum3,Liouville,Liouville2,SYK new,SYK new2,SYK new3,SYK new4,quench1} and gravitational physics\cite{Comments,bulk Yang,bulk spectrum Polchinski,bulk2,bulk3,bulk4,bulk5,syk-bh,SYK g new1,SYK g new2,SYK g new3,new g,new g2}. The SYK$_q$ model describes $N$ Majorana modes in 0+1-$d$ interacting randomly via $q$-fermion interactions \cite{Comments}. For simplicity, we focus on $q=4$ case and the Hamiltonian is then given by: 
\begin{align}
H_{\text{SYK}}[J_{i_1i_2i_3i_4},\chi]=\sum_{i_1i_2i_3i_4}\frac{J_{i_1i_2i_3i_4}}{4!}\chi_{i_1}\chi_{i_2}\chi_{i_3}\chi_{i_4}. \label{1}
\end{align}
Here $i_1,i_2...i_4=1, 2...N$ labels different modes of Majorana fermions. $J_{i_1i_2i_3i_4}$ are independent random Gaussian variables with $\overline{J_{i_1i_2i_3i_4}}=0$ and $\overline{J_{i_1i_2i_3i_4}^2}=3!J^2/N^3$. The model can be solved in the $1/N$ expansion and the two point correlation function is determined consistently by the Schwinger-Dyson equation with melon diagrams \cite{Comments}. In the low-temperature limit $\beta J \gg 1$, the system is found to be a strongly correlated non-Fermi liquid with low-energy holographic description \cite{bulk Yang}. In this system, without a spatial dimension, the Kadanoff-Baym equation can be solved efficiently in numerics, leading to exact quantum dynamics \cite{Sachdev,num2,num3}.

In this paper, we would like to study the dynamics of SYK model when coupled to an external bath with lower initial temperature. However, a general evolving bath requires a large amount of computational resource, which may make the problem intractable. To simplify the problem, following the idea of \cite{Yiming}, we model the bath also by an SYK model with a larger number of modes. We analyze the physical consequence of different coupling terms between two systems, including marginal coupling, irrelevant coupling, and relevant coupling \cite{yyz condensation,Balents,our,Altman}. We find in all cases, the energy firstly increases in time before decreasing, as expected from perturbative calculation \cite{Swingle}. For the marginal and the irrelevant coupling, the energy and the effective temperature of the system then relax to the thermal equilibrium monotonically with a different rate. While for the relevant case, both energy and effective temperature show oscillations. We also study the coupling and temperature dependence of different processes.

The paper is organized as follows: In section II, we describe our model and analyze its behavior for different couplings in thermal equilibrium. We describe the Keldysh path-integral used for calculating the quantum dynamics of our model in In section III. We then show numerical results in section IV for the marginal coupling case and section V for the irrelevant or relevant coupling case.

\section{The Model in Thermal Equilibrium}

\begin{figure}[t]
 	\center
 	\includegraphics[width=1\columnwidth]{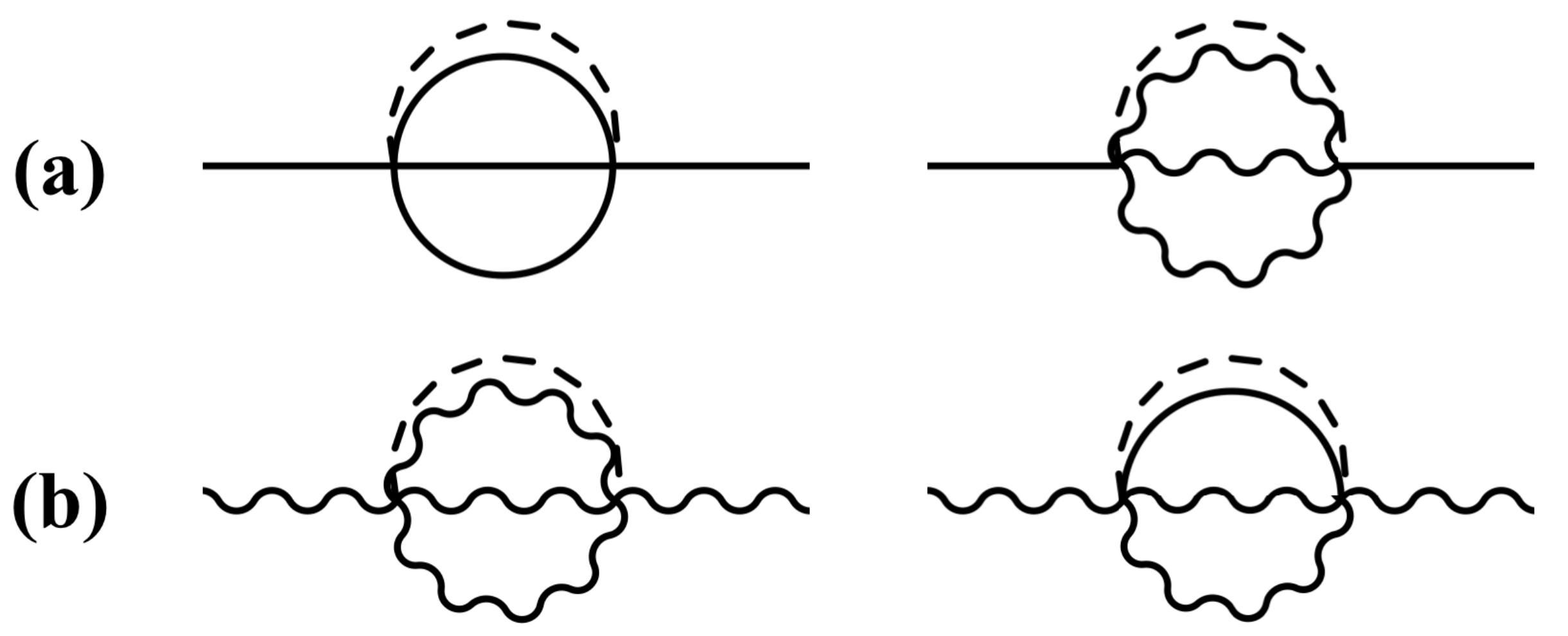}
 	\caption{The self-energy melon diagrams of the small SYK$_\chi$ coupled to SYK$_\psi$ bath for $n=3$. The solid line represents the Green's function of the $\chi$ fermion and the wavy line represents the Green's function of the $\psi$ fermion. The first three diagrams are of the order $N^0$ and the last diagram is proportional to $1/N$.} \label{fig} 	
 \end{figure}

As explained in the introduction, the model we considered in this paper is written as:
\begin{align}
&H=H_{\text{SYK}}[J_{i_1i_2i_3i_4},\chi]+H_{\text{SYK}}[\tilde{J}_{i_1i_2i_3i_4},\psi]\notag\\&\ \ \ +\sum_{ai_1i_2...i_n}\frac{V_{ai_1i_2...i_n}}{n!}\chi_a\psi_{i_1}\psi_{i_2}...\psi_{i_n}.\label{H}
\end{align}
where $H_{\text{SYK}}[J_{i_1i_2i_3i_4},\chi/\psi]$ is the standard SYK$_4$ Hamiltonian Eq. \eqref{1}. We choose the anti-commutation relation $\{\chi_i,\chi_j\}=\delta_{ij}$ and $\{\psi_i,\psi_j\}=\delta_{ij}$. $\chi$ is a small system with $N$ Majorana fermions and $\psi$ is a large system with $N^2$ fermions, which would be considered as a thermal bath of the small system. For each system there is an SYK$_4$ random interaction $J_{i_1i_2i_3i_4}$ or $\tilde{J}_{i_1i_2i_3i_4}$. We then randomly couple two systems by $V_{ai_1i_2...i_n}$ where $n$ is an odd number. All random interaction strength is assumed to be independent Gaussian variables with expectation and variance given by:
\begin{align}
&\overline{J_{i_1i_2i_3i_4}}=0,\ \ \ \ \ \ \overline{\tilde{J}_{i_1i_2i_3i_4}}=0,\ \ \ \ \ \ \overline{V_{ai_1i_2...i_n}}=0,\\
&\overline{J_{i_1i_2i_3i_4}^2}=\frac{3!J^2}{N^3},\ \ \ \overline{\tilde{J}_{i_1i_2i_3i_4}^2}=\frac{3!J^2}{N^3},\ \ \ \overline{V_{ai_1i_2...i_n}^2}=\frac{n!V^2}{N^{2n}}.
\end{align}
Here the numerical coefficient is chosen to cancel additional factors in melon diagrams. The power of $N$ is tuned to result in a well-defined non-trivial large-N theory, which is easiest to see by considering the self-energy of two-point correlators. We have shown the self-energy melon diagrams in Figure \ref{fig} for the $n=3$ case as an example. By straightforward counting, we could show that both diagrams in (a) and the first diagram in (b) is of the order $N^0$, while the last diagram in Figure \ref{fig} is proportional to $1/N$ \cite{Yiming}. As a result, although the small system $\chi$ is affected by the coupling $V$, the large system $\psi$ can still be approximated as isolated. This supports our identification of the large system by a thermal bath. 

Similar analysis also works for general $n$. As a result, the self-consistent equation for the two-point function is given by
\begin{align}
G^{-1}_\chi(\omega_n)&=-i\omega_n-\Sigma_\chi(\omega_n),\\ \Sigma_\chi(\tau)&=J^2G_\chi(\tau)^3+V^2G_\psi(\tau)^n,\label{Gchi}
\end{align}
for the $\chi$ fermions and
\begin{align}
G^{-1}_\psi(\omega_n)&=-i\omega_n-\Sigma_\psi(\omega_n),\\ \Sigma_\psi(\tau)&=J^2G_\psi(\tau)^3. \label{Gpsi}
\end{align}
for $\psi$ fermions. Here we have defined $G_\chi(\tau)=\left<T_\tau\chi_i(\tau)\chi_i(0)\right>$ and $G_\psi(\tau)=\left<T_\tau\psi_i(\tau)\psi_i(0)\right>$, with $T_\tau$ being the time-ordering operator in imaginary time. 

From Eq.\eqref{Gpsi}, the Green's function of the large system is the same as a single SYK$_4$ model: the scaling dimension of $\psi$ is $\left[\psi\right]=1/4$. At the zero-temperature limit, this leads to:
\begin{align}
G_\psi(\tau)=b_\psi\frac{\text{sgn}(\tau)}{|\tau|^{1/2}}, \ \ \ \ \ \ 4\pi J^2b_\psi^4=1，
\end{align}
and the Green's function at finite temperature is then given by conformal mapping $\tau=\tan\frac{\pi\tau'}{\beta}$\cite{Comments}. On the other hand, due to the competition of the two terms in \eqref{Gchi}, the physics of small system $\chi$ is very different for $n=1$, $n=3$ and $n> 3$.

(1). For the $n=3$ case, the coupling term $V_{ai_1i_2i_3}$ shares the same scaling dimension with the on-site SYK$_4$ interaction. As a result, the scaling dimension of $\chi$ is also $1/4$. Nevertheless, the interaction would renormalize the coefficient of the Green's function: 
\begin{align}
G_\chi(\tau)=b_\chi\frac{\text{sgn}(\tau)}{|\tau|^{1/2}}, \ \ \ \ \ \ 4\pi J^2b_\chi^4+4\pi V^2b_\chi b_\psi^3=1.
\end{align}
We could define $b_\chi=\eta b_\psi$, and then we have $\eta^4+\frac{V^2}{J^2}\eta=1$. 

The thermalization of a quantum system is closely related to the information scrambling \cite{upper Hartnoll,lower Hartnoll,upper Lucas,lower Blake}. As a result, it is useful to compute the Lyapunov exponent for this sytem. The out-of-time order correlation function at inverse temperature $\beta$ is defined as $$F_\chi(t_1,t_2)=\left<\chi_i(t_1-i\beta/2)\chi_j(-i\beta/2)\chi_i(t_2)\chi_j(0)\right>.$$ To the leading order of $1/N$, the self-consistent equation for $F_\chi(t_1,t_2)$ in long-time limit can be written as:
\begin{align}
F_\chi(t_1,t_2)&=\int dt_3dt_4 K_R(t_1,t_2;t_3,t_4)F_\chi(t_3,t_4),\\
 K_R(t_1,t_2;t_3,t_4)&=-3J^2G_{R,\chi}(t_{13})G_{R,\chi}(t_{24})G_{W,\chi}(t_{34})^2.
\end{align}
Where $G_{R,\chi}(t)=-i\theta(t)\left<\{\chi_i(t),\chi_i(0)\}\right>$ is the standard retarded Green's function and $G_{W,\chi}(t)\equiv\left<\chi(t-i\beta/2)\chi(0)\right>$. For simplicity, we take $\beta=2\pi$. Using the assumption:
\begin{align}
F_\chi(t_1,t_2)=\frac{e^{-h\frac{t_1+t_2}{2}}}{\left(\cosh\frac{\pi(t_1-t_2)}{\beta}\right)^{1/2-h}},
\end{align}
it can be shown that the self-consistent equation is satisfied if $3\eta^4=1-2h$. This gives a Lyapunov exponent
\begin{align}
\lambda_L=\frac{3\eta^4-1}{2}\approx1-\frac{3V^2}{8J^2}+O\left(\left(\frac{V^2}{J^2}\right)^2\right).
\end{align}
Physically, due to the marginal coupling, information leaks from the $\chi$ system into the thermal bath $\psi$. As a result, the scrambling of information in the small system becomes slower \cite{Yiming}.

(2). For the $n=1$ case, the inter-site coupling is relevant near the original SYK$_4$ fixed point and the in the zero-temperature limit, the $\chi$ system is driven into a new phase with scaling dimension $[\chi]=1-[\psi]$:
\begin{align}
G_\chi(\tau)=b_\chi\frac{\text{sgn}(\tau)}{|\tau|^{3/2}}, \ \ \ \ \ \ 4\pi V^2b_\chi b_\psi=1.
\end{align}
Here we neglected the contribution from $J_{i_1i_2i_3i_4}$ since it is irrelevant near this new fixed point. After Fourier transformation, this shows the spectral function vanishes as $\sqrt{\omega}$ for small $\omega$, indicating the system is non-chaotic. One could take into account the contribution of $\partial_\tau$ term, and the Green's function then has the form of $G_\chi^{-1}(\omega)\sim \omega+V^2/\sqrt{J \omega}$, which gives a minimal at $\omega_0 \sim J(V^2/J^2)^{2/3}$. This gives an approximation for the peak of the spectral function.

For the out-of-time order correlation function $F_\chi(t_1,t_2)$, by counting the $N$ factor, one could find the random coupling $V_{ai_1}$ gives no contribution to $F_\chi(t)$ to the $1/N$ order in the conformal limit (which is also true for general $n$ \cite{Yiming}), as a result we have $\lambda_L \beta\rightarrow 0$ as $\beta \rightarrow \infty$. This system is non-chaotic in the low-energy limit.

(3). Finally, for the $n>3$ case, the coupling term is irrelevant near the decoupled SYK$_4$ fixed point. As a result, we have:
\begin{align}
G_\chi(\tau)=G_\psi(\tau).
\end{align}
In the low-temperature $\beta J \rightarrow \infty$ we could still have a non-Fermi liquid $\chi$ with maximal chaos $\beta \lambda_L\rightarrow 2\pi$.

\section{Evaporation Dynamics on Keldysh Contour}
Different thermal behaviors for systems with different $n$ indicates they should also have different quench dynamics. In this work we focus on such evaporation process by preparing an initial thermal ensemble with $V_{ai_1i_2...i_n}=0$ at $t<0$, and turn on the interaction $V_{ai_1i_2...i_n}$ at $t=0$. This quench problem can be analyzed on the Keldysh contour \cite{Kamenev}, where fields $\chi_+$, $\psi_+$ live on the upper ($+$) contour while $\chi_-$, $\psi_-$ live on the lower ($-$) contour. The partition function on Keldysh contour then is given by:
\begin{align}
\mathcal{Z}&=\int d J d\tilde{J}dVP(J,\tilde{J},V)\mathcal{D}\chi_+\mathcal{D}\psi_+\mathcal{D}\chi_-\mathcal{D}\psi_-e^{i\int dt L},\\
L&=\sum_i\frac{1}{2}\chi_{i,\alpha}(\hat{G}^0)^{-1}_{\alpha\beta}\chi_{i,\beta}+\sum_i\frac{1}{2}\psi_{i,\alpha}(\hat{G}^0)^{-1}_{\alpha\beta}\psi_{i,\beta}\notag\\&\ \ -H[\chi_+,\psi_+]+H[\chi_-,\psi_-]. \label{L}
\end{align}
Here $\alpha$, $\beta=\pm$ and $P(J,\tilde{J},V)$ is Gaussian the distribution function for random variables. On such contour, the Green's function $\hat{G}$ is defined as
\begin{align}
\hat{G}_{\chi,\alpha\beta}(t,t')=-i\left<\chi_\alpha(t)\chi_\beta(t')\right>=\begin{pmatrix}
G^{T}_\chi(t,t')&G^{<}_\chi(t,t')\\
G^{>}_\chi(t,t')&G^{\tilde{T}}_\chi(t,t')
\end{pmatrix}.
\end{align}
And similar definition works for $\hat{G}_\psi$. We have $(\hat{G}^0)=(\hat{G})_{J,V=0}$ is the non-interacting limit of the Green's function \cite{Kamenev}. For Majorana fermions, we have the relation $G^>(t,t')=(G^<(t,t'))^*$. Green's functions in this $\pm$ basis are related to the retarded, advanced, Keldysh components of the Green's function by Keldysh rotation:
\begin{align}
G_{R}(t,t')&=\theta(t-t')(G^>(t,t')-G^<(t,t')),\label{GR}\\
G_{A}(t,t')&=\theta(t'-t)(G^<(t,t')-G^>(t,t')),\\
G_{K}(t,t')&=G^<(t,t')+G^>(t,t').\label{GK}
\end{align}
$H[\chi_\pm,\psi_\pm]$ is defined by replacing the operator $\chi_i$ by corresponding field $\chi_{i,\pm}$, with an additional $\theta(t)$ factor in the inter-site coupling:
\begin{align}
H[\chi_\pm,\psi_\pm]=&\sum_{i_1i_2i_3i_4}\frac{J_{i_1i_2i_3i_4}}{4!}\chi_{i_1,\pm}\chi_{i_2,\pm}\chi_{i_3,\pm}\chi_{i_4,\pm}\notag\\&+\sum_{i_1i_2i_3i_4}\frac{\tilde{J}_{i_1i_2i_3i_4}}{4!}\psi_{i_1,\pm}\psi_{i_2,\pm}\psi_{i_3,\pm}\psi_{i_4,\pm}\notag\\&+\theta(t)\sum_{ai_1...i_n}\frac{V_{ai_1i_2...i_n}}{n!}\chi_{a,\pm}\psi_{i_1,\pm}...\psi_{i_n,\pm}.
\end{align} 
The Schwinger-Dyson equation for two-point correlators contains the same melon diagrams with the imaginary time calculation shown in Figure. \ref{fig}. This gives the self-energy:
\begin{align}
\hat\Sigma_{\chi,\alpha\beta}(t,t')&\equiv\begin{pmatrix}
\Sigma_\chi^{T}(t,t')&-\Sigma_\chi^{<}(t,t')\\
-\Sigma_\chi^{>}(t,t')&\Sigma_\chi^{\tilde{T}}(t,t')
\end{pmatrix}_{\alpha\beta}\notag\\&=-J^2\alpha \beta G^3_{\chi,\alpha\beta}(t,t')\notag \\ &\ \ \ -V^2\alpha \beta (-1)^{\frac{n+1}{2}}\theta(t)\theta(t')G^{n}_{\psi,\alpha\beta}(t,t'),\\
\hat\Sigma_{\psi,\alpha\beta}(t,t')&\equiv\begin{pmatrix}
\Sigma_\psi^{T}(t,t')&-\Sigma_\psi^{<}(t,t')\\
-\Sigma_\psi^{>}(t,t')&\Sigma_\psi^{\tilde{T}}(t,t')
\end{pmatrix}_{\alpha\beta}\notag \\ &=-J^2\alpha \beta G^3_{\psi,\alpha\beta}(t,t').
\end{align}
Similarly to the Green's function, we could also define the retarded, advanced, Keldysh components of the self-energy for both $\chi$ and $\psi$ as:
\begin{align}
\Sigma_{R}(t,t')&=\theta(t-t')(\Sigma^>(t,t')-\Sigma^<(t,t')),\\
\Sigma_{A}(t,t')&=\theta(t'-t)(\Sigma^<(t,t')-\Sigma^>(t,t')),\\
\Sigma_{K}(t,t')&=\Sigma^<(t,t')+\Sigma^>(t,t').
\end{align}

As in the imaginary-time calculation, the bath $\psi$ is not affected by the small system. As a result, we know $G_\psi$ is always given by the equilibrium result. The spectral function $A_\psi(\omega)=-\frac{1}{\pi}\text{Im}G_{R,\psi}(\omega)$ in thermal equilibrium with temperature $T_\psi$ can be determined numerically by the self-consistent equation of retarded Green's function:
\begin{align}
G_{R,\psi}(\omega)^{-1}&=\omega-\Sigma_{R,\psi}(\omega), \label{GR1}\\
\Sigma_{R,\psi}(\omega)&=-iJ^2\int_0^\infty dt e^{i\omega t}(n_\psi(t)^3+(n_\psi(t)^*)^3), \label{GR2}\\
n_\psi(t)&=\int d\omega e^{-i\omega t} A_\psi(\omega)n_F(\omega, T_\psi), \label{GR3}
\end{align}
where $n_F(\omega, T_\psi)$ is the Fermi-Dirac distribution function at temperature $T_\psi$ and we have used the relation \cite{Kamenev}:
\begin{align}
G^>(\omega)=-in_F(-\omega,T)A(\omega),
\end{align}
valid for Majorana fermions on thermal equilibrium.

In contrast, the small $\chi$ system is driven by its coupling to the large system and becomes time-dependent. For such an evolution problem, it is better to write the self-consistent equation in the form of Kadanoff-Baym equations in real-time for $t>0$ using the Langreth rules \cite{book}, this gives:
 \begin{align}
 i\partial_{t_1}G^>_\chi(t_1,t_2)=\int d t_3 (&\Sigma^R_\chi(t_1,t_3)G^>_\chi(t_3,t_2)\notag \\ &+\Sigma^>_\chi(t_1,t_3)G^A_\chi(t_3,t_2)), \label{eq1}\\
 -i\partial_{t_2}G^>_\chi(t_1,t_2)=\int d t_3 (&G^R_\chi(t_1,t_3)\Sigma^>_\chi(t_3,t_2)\notag \\ &+G^>_\chi(t_1,t_3)\Sigma^A_\chi(t_3,t_2)).\label{eq2}
 \end{align}
 In these equations, the evolution of $G_\chi^>(t,t')$ only depends on information of $G_\chi^>(t_1,t_2)$ with $t_1<t$ and $t_2<t'$, which make the causal structure explicit. The initial condition of $G_\chi^>(t,t')$ is given by the thermal solution:
\begin{align}
G_\chi^>(t,t')=G_\chi^>(t-t'),\ \ \ \ \text{for}\ \ t,t'<0. \label{ini}
\end{align}
where $G_\chi^>(t-t')$ is determined similar to \eqref{GR1}, \eqref{GR2} and \eqref{GR3}, with $T_\psi$ replaced by $T_\chi$. Solving \eqref{eq1} and \eqref{eq2} with initial condition \eqref{ini} leads to exact (for large $N\gg1$) quench dynamics of the small SYK model when coupled to a large SYK bath. We have checked that if $V=0$, the numerical evolution preserve the translation symmetry $G_\chi^>(t,t')=G_\chi^>(t-t')$.
\begin{figure}[t]
 	\center
 	\includegraphics[width=1\columnwidth]{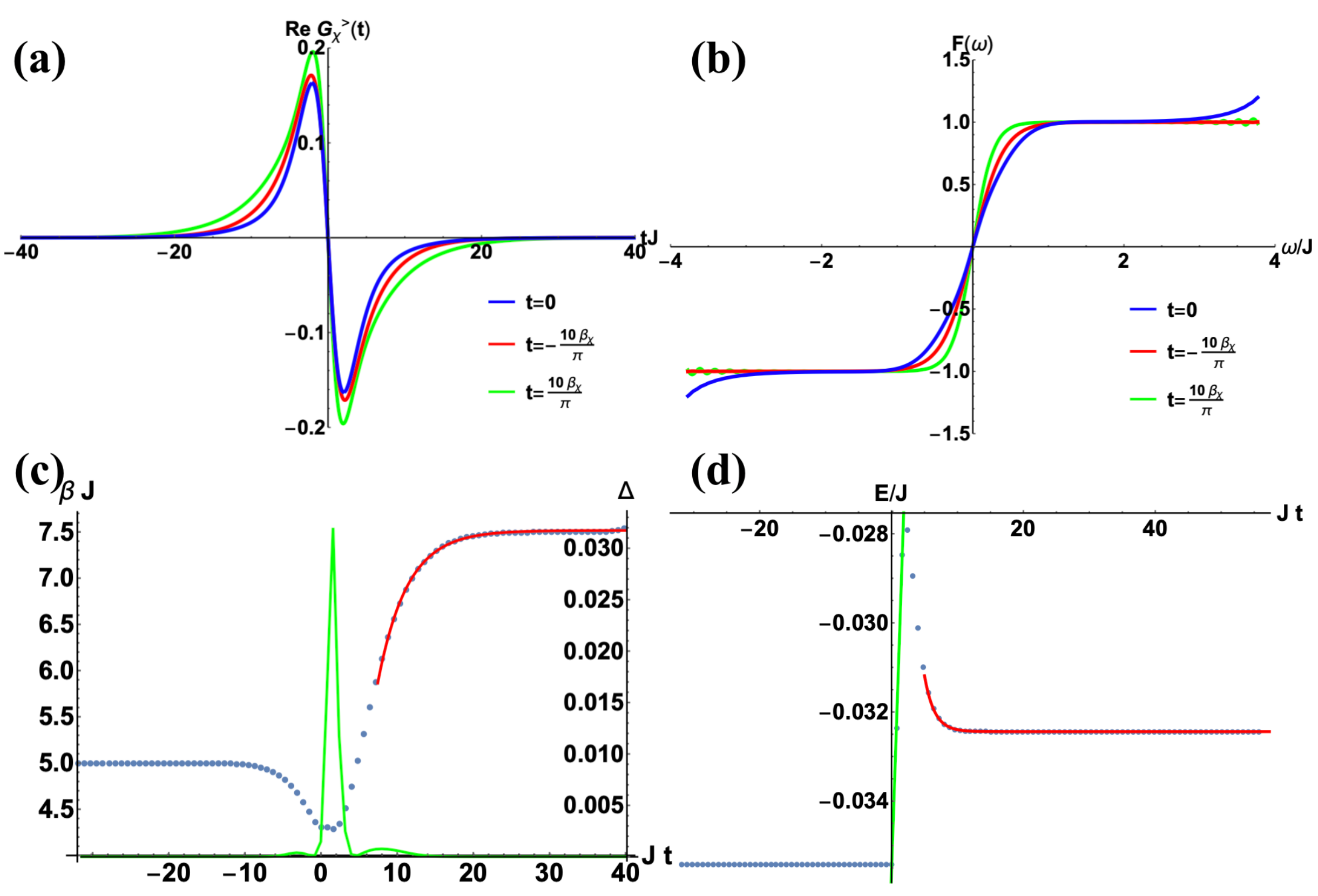}
 	\caption{The result of quench dynamics for $n=3$ with $V/J=0.6$, $T_\chi=0.2 J$ and $T_\chi=1.5 T_\psi$. (a). The real part of $G^>_\chi(t+\frac{t_r}{2},t-\frac{t_r}{2})$ as a function of $t_r$ for different $t$. (b). $F(\omega,t)=G_{\chi,K}(\omega,t)/\left(G_{\chi,R}(\omega,t)-G_{\chi,A}(\omega,t)\right)$ for different time $t$. (c). The evolution of effective temperature $T(t)$. The red line shows the result of exponential fitting of the late-time behavior. The green line represents the distance between $F(\omega)$ and $1-2n_F(\omega,T(t))$ defined by \eqref{Delta}, we take the cutoff $\Lambda$ by requiring $F(\Lambda)=0.8$. (d). The evolution of energy $E(t)$ determined by \eqref{ene}. The green line is a fit for the short-time linear increase of energy and the red line is a late-time exponential fit for the relaxation of energy.} \label{fig2}
 \end{figure}

After numerical evolution, we define the effective temperature at time $t$:
\begin{align}
1/T(t)&=2\frac{d}{d\omega}\left(\frac{G_{\chi,K}(\omega,t)}{G_{\chi,R}(\omega,t)-G_{\chi,A}(\omega,t)}\right)_{\omega=0}\notag\\&\equiv2\frac{d}{d\omega}\left(F(\omega,t)\right)_{\omega=0}.
\end{align}
Here we have performed the Wigner transformation of Green's functions:
\begin{align}
G(\omega,t)=\int dt' e^{i\omega t'}G(t+\frac{t'}{2},t-\frac{t'}{2}).
\end{align}
We also define 
\begin{align}
\Delta=\int_{|\omega|<\Lambda} d\omega \left(\frac{G_{\chi,K}(\omega,t)}{G_{\chi,R}(\omega,t)-G_{\chi,A}(\omega,t)}-(1-2n_F(T))\right)^2,\label{Delta}
\end{align}
which characterize the difference between the numerical result and a thermal distribution function in low-energy limit with cutoff $\Lambda$. We could also define instantaneous energy of the $\chi$ system by 
\begin{align}
E(t_0)=\sum_{i_1i_2i_3i_4}\frac{1}{4!}\overline{\left<J_{i_1i_2i_3i_4}\chi_{i_1}(t_0)\chi_{i_2}(t_0)\chi_{i_3}(t_0)\chi_{i_4}(t_0)\right>}\label{ene}
\end{align}
To express this formula in terms of $G^>_\chi$, we add an source term to the Lagrangian in Eq. \eqref{L}:
\begin{align}
\Delta L(t_0)=\sum_{i_1i_2i_3i_4}\frac{J_{i_1i_2i_3i_4}a\delta(t-t_0)}{4!}\chi_{i_1,+}\chi_{i_2,+}\chi_{i_3,+}\chi_{i_4,+}
\end{align}
Then by taking derivative to the standard $G-\Sigma$ action \cite{Comments}, it is straightforward to prove the relation:
\begin{align}
E(t_0)&=\left(\frac{d \ln \mathcal{Z}(a)}{da}\right)_{a\rightarrow0}\notag\\&=i\frac{J^2}{4}\int^{t_0} dt \left(G^>_\chi(t_0,t)^4-G^<_\chi(t_0,t)^4\right).\label{E0for}
\end{align}

\section{Marginal Coupling: Energy Increase and Relaxation}
We first consider the $n=3$ case where both on-site interaction $J$ and coupling to bath $V$ contribute to the low-energy physics. We take $J/T_\chi=5$ throughout the numerics.

In Figure \ref{fig2}, we have shown a typical numerical result. We choose the parameter to be $V/J=0.6$ and $T_\chi=1.5 T_\psi$. In (a), we plot $G^>_\chi(t+\frac{t_r}{2},t-\frac{t_r}{2})$ for different time $t=0,\ \pm 10 \beta_\chi/\pi$. Based on this results, we could compute the distribution $F(\omega,t)$ at different time, as shown in (b). From the result of $F(\omega,t)$, we could determine the effective temperature $T(t)=1/\beta(t)$ in (c). We also check the distance between $F(\omega)$ and $1-2n_F(\omega,T(t))$ (the green curve), where we take the cutoff $\Lambda$ satisfies $F(\Lambda)=0.8$. This result suggests the low-energy behavior of the system can be approximated by an thermal ensemble for almost any $t$. In Figure (d), we show the energy $E(t)$ of the $\chi$ system. For short time limit, the system absorbs energy from the coupling linearly (fitted by the green line) and in long-time limit the energy flows into the bath. 

\begin{figure}[t]
 	\center
 	\includegraphics[width=1\columnwidth]{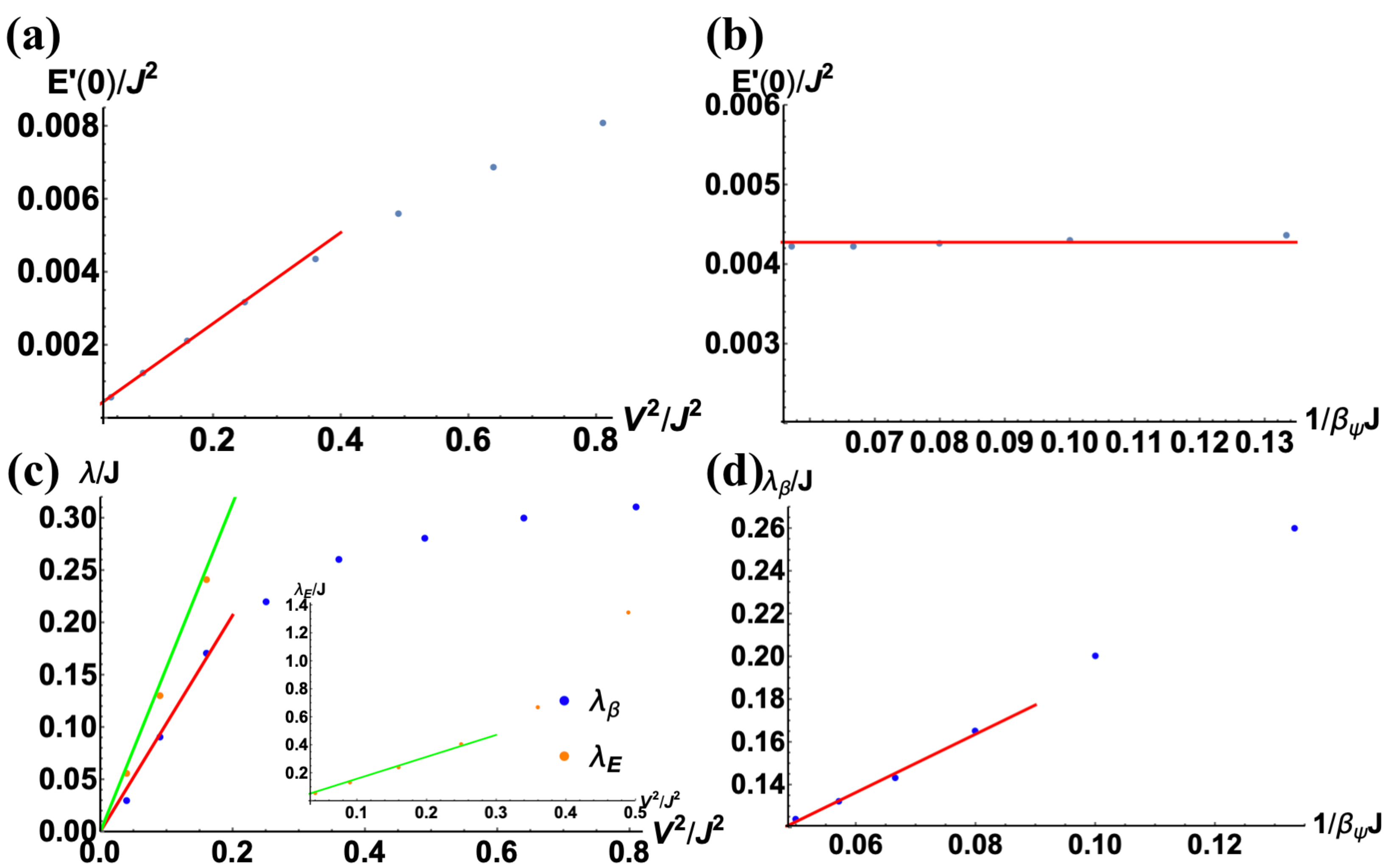}
 	\caption{(a). The energy absorption rate $E'(0)/J^2$ as a function of $V/J$ for $T_\chi=0.2 J$ and $T_\chi=1.5 T_\psi$. The red line is a linear fit for the first four points. (b). The energy absorption rate $E'(0)/J^2$ as a function of $T_\psi/J$ for $T_\chi=0.2 J$ and $V/J=0.6$. There is almost no temperature dependence. (c). Relaxation rate $\lambda_\beta$ and $\lambda_E$ as a function of $V/J$ for $T_\chi=0.2 J$ and $T_\chi=1.5 T_\psi$. The red and the green line is a linear fit for the first three points. (d). Relaxation rate $\lambda_\beta$ as a function of $T_\psi/J$ for $T_\chi=0.2 J$ and $V/J=0.6$. The red line is a linear fit for the first four points.} \label{fig3}
 \end{figure}
Since the system is a many-body chaotic non-Fermi liquid, we expect the effective temperature should show fast relaxation to the thermal equilibrium with $T(\infty)=T_\psi$. However, interestingly we find the effective temperature will increase first before it starts to decrease, which is clear from Figure \ref{fig2} (c) and (d). Physically, this is because when we quench the system by adding new interaction term, we create some excitations in the system first. Only then the energy of the small system begins to dissipate into the thermal bath, and the system is cooled down. This phenomenon is firstly discussed by Swingle in talk \cite{Swingle} for general quantum systems, where a perturbative calculation for the system-bath coupling, as well as exact diagonalization for SYK models coupled to a wire bath, have already been worked out. It is also found to be related to the averaged null energy condition in holographic systems \cite{holo}. Here we find that such effect also holds for finite coupling strength in our system.

We quantify this temperature increase by studying the behavior of $E'(0^+),$ as a function of $V/J$ and $T_\psi/J$. For simplicity, we would drop the $+$ sign later. Physically, increasing $V/J$ would excite more excitations and as a result, $E'(0)$ should become larger. For small $V$, perturbatively we have
\begin{align}
\frac{dE}{dt}&=\sum\left<\left[\frac{J_{i_1i_2i_3i_4}}{4!}\chi_{i_1}\chi_{i_2}\chi_{i_3}\chi_{i_4},\frac{V_{ai_1i_2i_3}}{3!}\chi_a\psi_{i_1}\psi_{i_2}\psi_{i_3} \right]\right>\notag \\
&\propto \sum\left<J_{i_1i_2i_3i_4}\chi_{i_1}\chi_{i_2}\chi_{i_3}V_{i_4j_1j_2j_3}\psi_{j_1}\psi_{j_2}\psi_{j_3}\right>\notag \\ &\sim J^2V^2 \int dt_1dt_2G_\chi^3(t_1)G_\chi(t_1-t_2)G_\psi^3(t_2+\epsilon)\notag\\
&\sim V^2 G_\psi^3(\epsilon)\propto V^2 \label{Vs}
\end{align}
 Where we have split the operators by cutoff $\epsilon\sim 1/J$ to avoid possible divergences in the third line. The Green's function represents either advanced or other components of the Green's function, whose specific choice is not important but could be determined using Eq. \eqref{eq1}, \eqref{eq2} and \eqref{E0for}. 

We indeed find such behavior in numerical results shown in Figure \ref{fig3} (a) and (b). In (a), we set $T_\chi=0.2 J$ and $T_\chi=1.5 T_\psi$. For small $V^2/J^2$, We find approximately $E'(0)\propto V^{2}$. On the other hand, the $E'(0)$ is also found to be almostly independent of $T_\psi/J$.

Then we consider the long-time limit where the energy of the small system $\chi$ finally decays into the bath. We could define two different relaxation rate: 

1. the relaxation of $\beta(t)$ defines $\lambda_\beta$ by $$\beta(t)\sim\beta_\psi-c_0\exp(-\lambda_\beta t),$$ which is similar to the thermalization rate $\Gamma$ for an isolated system \cite{Sachdev}. 

2. The relaxation of energy $E(t)$, given by $$E(t)\sim E(\infty)-c_0'\exp(-\lambda_E t).$$ 

From Figure. \ref{fig3} (c), we see $\lambda_E<\lambda_\beta$, which is reasonable since the relaxation of energy is a necessary condition for thermal equilibrium. 

\begin{figure}[t]
 	\center
 	\includegraphics[width=1\columnwidth]{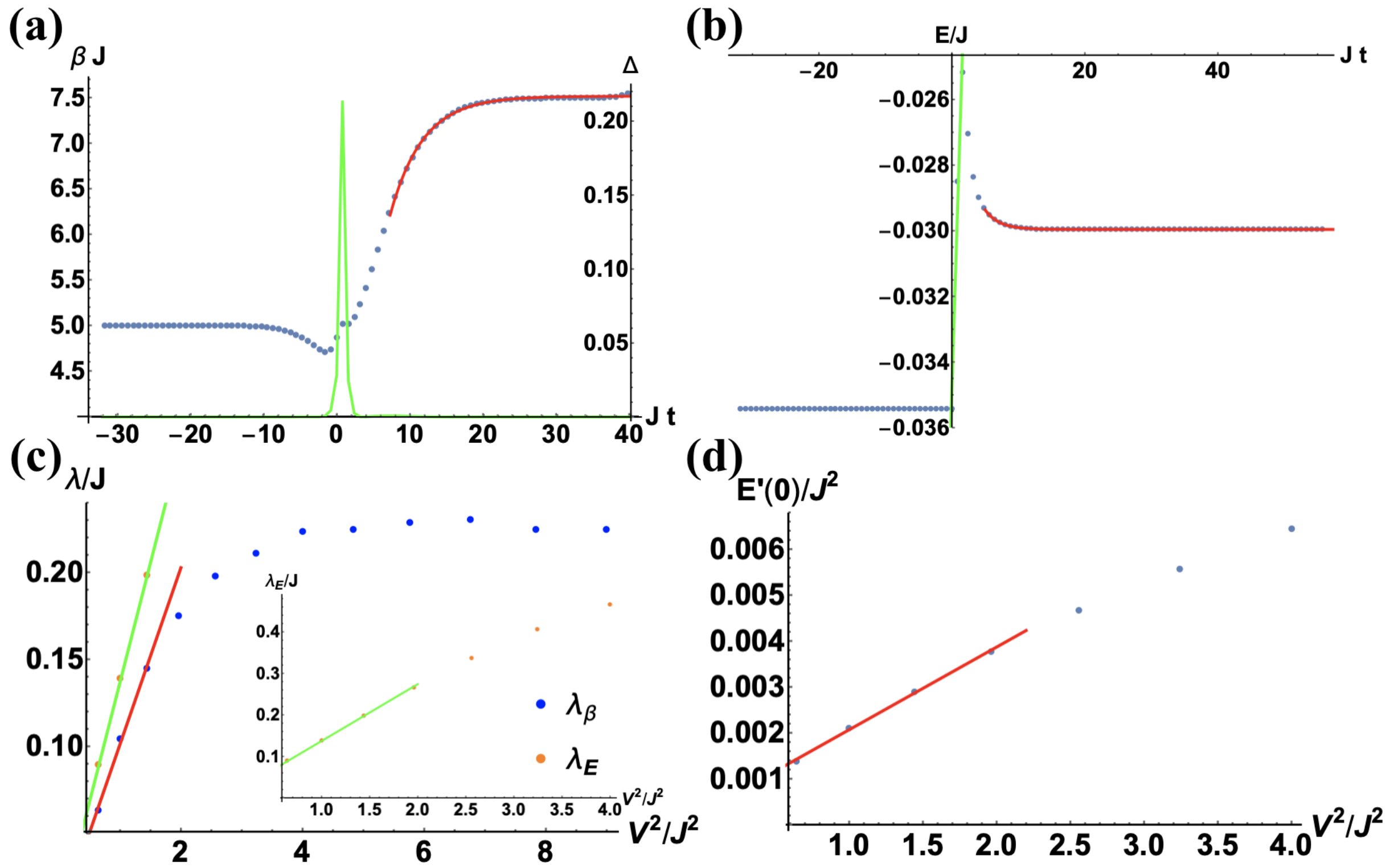}
 	\caption{The quench dynamics for $n=5$. (a-b). The result of quench dynamics with $V/J=2$, $T_\chi=0.2 J$ and $T_\chi=1.5 T_\psi$. In (a), we plot evolution of effective temperature $\beta(t)$ and the distance $\Delta$ (Again, we take $F(\Lambda)=0.8$.). The red line shows the result of the exponential fitting of the late-time behavior. (b). The evolution of energy $E(t)$. The green line is a fit for the short-time linear increase of energy and the red line is a late-time exponential fit for the relaxation of energy. (c). Relaxation rate $\lambda_\beta$ and $\lambda_E$ as a function of $V/J$ for $T_\chi=0.2 J$ and $T_\chi=1.5 T_\psi$. The red and the green line is a linear fit for the first three points. (d). The energy absorption rate $E'(0)/J^2$ as a function of $V/J$ for $T_\chi=0.2 J$ and $T_\chi=1.5 T_\psi$. The red line is a linear fit for the first four points.} \label{fig4}
 \end{figure}
 For small $V/J$, perturbatively, we expect the energy relaxation rate to be proportional to $V^2$, which is consistent with the time-scale that this system becomes thermal, as shown in Figure \ref{fig3} (c). Physically, here the process is dominated by the energy relaxation from the $\chi$ system into the bath $\lambda_\beta \sim \lambda_E$. However, if we further increase the inter-site coupling or go to the low-temperature limit, the relaxation rate would saturate to $\lambda \propto T_\psi$. This is because the energy flows into the bath quicker than the system itself becomes thermalized. As a result, we expect $\lambda_\beta \sim \Gamma\ll \lambda_E$, where $\Gamma$ is known to be bounded by temperature $ T_\psi$ \cite{Hqm}. 
\section{Irrelevant and Relevant Coupling}

In this section we consider different coupling terms with $n=5$ and $n=1$. As discussed in previous sections, they correspond to the irrelevant and relevant coupling case. 

We firstly consider the irrelevant case $n=5$. Since the coupling is irrelevant, it should have a neglectable effect in the zero-temperature limit. While for the finite temperature case, we expect the influence of the coupling on the system is much smaller than the $n=3$ case and the system should evolve adiabatically even for large $V/J>1$, as shown in Figure \ref{fig4} (a) where $V/J=2$. Nevertheless, we find qualitatively similar behavior compared to the $n=3$ case, where the system ultimately relax to a thermal ensemble with $T(\infty)=T_\psi$. The figure (b) shows the system also absorb energy first with almost constant rate and then the energy flow back to the bath.

\begin{figure}[t]
 	\center
 	\includegraphics[width=1\columnwidth]{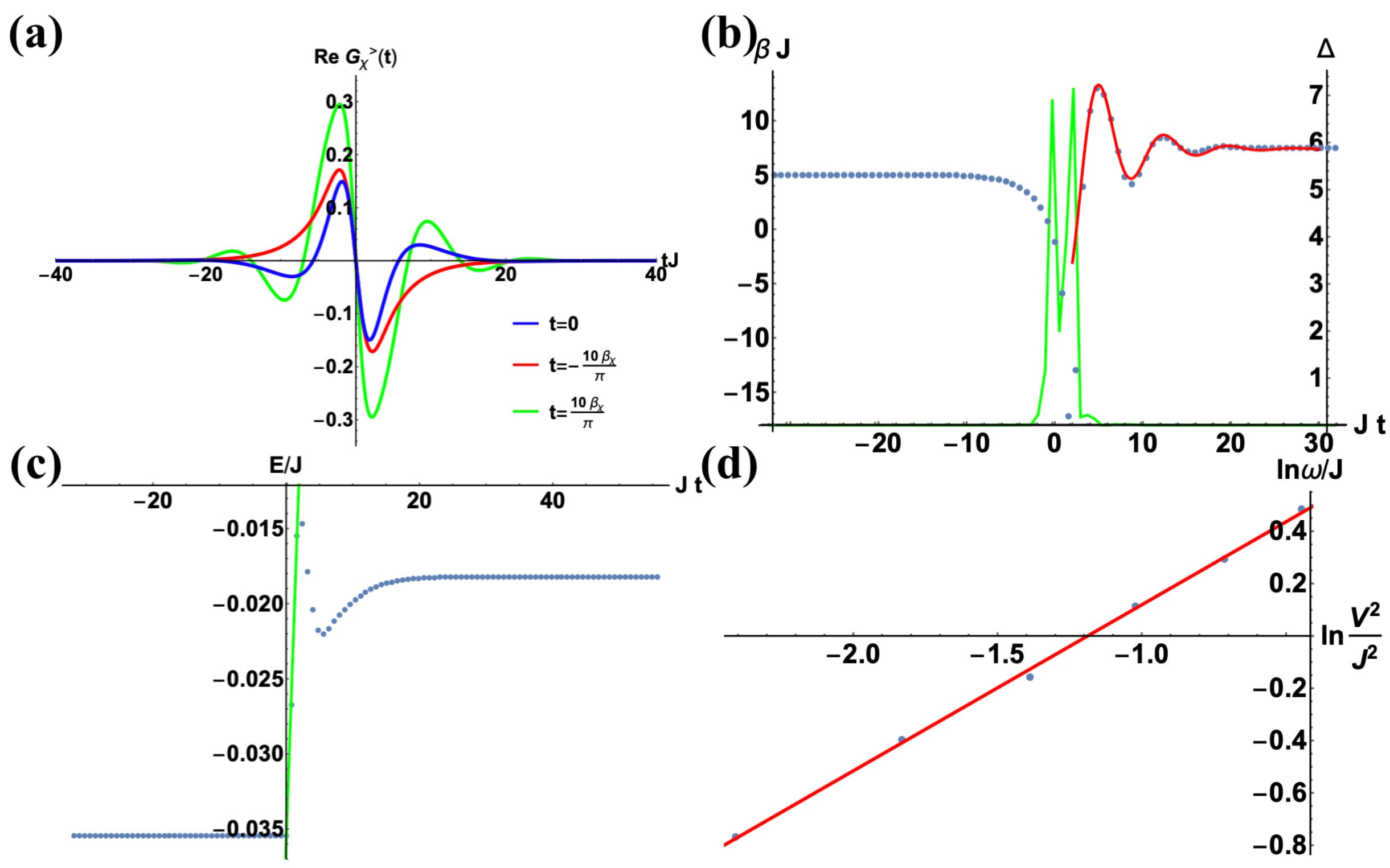}
 	\caption{The quench dynamics for $n=1$. (a-c). The result of quench dynamics with $V/J=0.5$ $T_\chi=0.2 J$ and $T_\chi=1.5 T_\psi$. (a). The real part of $G^>_\chi(t+\frac{t_r}{2},t-\frac{t_r}{2})$ as a function of $t_r$ for different $t$. In (b), we plot evolution of effective temperature $\beta(t)$ and the distance $\Delta$ (Again, we take $F(\Lambda)=0.8$.). The red line shows the result for fitting of the late-time behavior. (c). The evolution of energy $E(t)$. The green line is a fit for the short-time linear increase of energy and the red line is a late-time exponential fit for the relaxation of energy. (d). Oscillation frequency $\omega$ as a function of $V^2/J^2$ for $T_\chi=0.2 J$ and $T_\chi=1.5 T_\psi$ in the $log$-$log$ plot. The red line is a linear fit suggesting $\log(\omega)\sim 0.63\log(V^2/J^2)+\text{cons.}$} \label{fig5}
 \end{figure}
 In figure (c), we further study the relaxation rate $\lambda_E$ and $\lambda_\beta$ of the system as a function of $V/J$. For the relaxation rate, we need a larger $V/J$ to make the energy absorption much quicker than the thermalization rate of the effective temperature. Nevertheless, when this occurs, we find the $\lambda_\beta$ is of the same order as the $n=3$ case as expected. We also check that this saturation value is proportional to $T_\psi$.

 In figure (d), we plot the energy absorption rate $\lambda_E'(0)$ as a function of $V/J$. Naturally, for small $V^2/J^2$, $\lambda_E'(0)\propto V^2/J^2$, which is determined from the perturbation theory. Similar to Eq. \eqref{Vs}, we should also expect $E'(0)$ would not have significant temperature dependence for small $V/J\ll 1$. 

On the contrary, the low energy behavior is different with or without coupling $V$ for the relevant case $n=1$. Consequently, the quench dynamics, in this case, is very different from previous results, as shown in Figure \ref{fig5}. Firstly, since the spectral function for $V\neq 0$ has a peak around $\pm \omega_0 \neq 0$, the Green's function in real-time $G^>_\chi$ show an oscillation in real-time for $t>0$, as shown in Figure \ref{fig5} (a). 

The fact that the collective mode has specific frequency is also reflected in the evolution of effective temperature $\beta(t)$, as shown in (b). To extract the oscillation frequency, we fit the late time behavior as:
\begin{align}
\beta(t)=c_0+d_0\exp(-\lambda t)\sin(\omega t+f_0).
\end{align}
The result of $\omega$ as a function of $V^2/J^2$ is shown in (d) for $T_\chi=0.2 J$ and $T_\chi=1.5 T_\psi$. The linear fit suggests $\log(\omega)\sim 0.63\log(V^2/J^2)+\text{cons.}$, which is close to the analytical approximation $\log(\omega_0)\sim0.67\log(V^2/J^2)+\text{cons.}$ We also find the energy of the system shows oscillations in (c).

\section{Summary and Outlook}
In this paper, we couple the small SYK system $\chi$ to a large SYK bath $\psi$ with lower temperature. In thermal equilibrium, depending on the number of $\psi$ operator $n$ in the coupling term, we find the system shows different behaviors: 

(1). For $n=3$, the coupling is marginal and the $\chi$ system is dressed by $\psi$, and the Lyapunov exponent deviates from maximal chaotic by a constant factor in the low-energy limit.

(2). For $n=1$, the coupling is relevant. The single-particle spectra show a peak at a finite frequency and the system is not chaotic in the low-temperature limit. 

(3). For $n=5$, the coupling is irrelevant and does not contribute to the low-energy limit.

Based on this knowledge, we further consider their quench dynamics. In all cases, the system firstly absorbs energy with almost constant energy absorption rate $E'(0)$ before the energy relaxes into the large bath. We find $E'(0)$ is determined by UV physics with almost no temperature dependence. 

(1). For both $n=3$ and $n=5$, the energy and the temperature decay monotonically for the relaxation process. For small system bath coupling $V$, we find both the temperature decay rate $\lambda_\beta$ and the energy decay rate $\lambda_E$ are proportional to $V^2/J^2$. While for $V^2/J^2 \gg 1$, the energy relaxes quickly and after that, the system gradually approaches the thermal equilibrium with $\lambda_\beta$ bounded by $1/\beta$.
For finite temperature, the main difference for the $n=3$ and $n=5$ case is that for $n=5$ we need much larger coupling $V$ to get a moderate relaxation rate. 

(2). The situation is very different for $n=1$. In this case, since the single-particle spectra show a peak at a finite frequency in the thermal ensemble, the relaxation is non-monotonically with oscillations in both temperature and energy. The oscillation frequency of the temperature is found to be the same as the quasi-particle energy.

It is interesting to generalize the set-up in this work to further study the evaporation dynamics. As an example, it would be interesting if one could study the evaporation of the coupled SYK model \cite{Xiaoliang}, where negative specific heat regime exists for intermediate temperature. This case would mimic the Hawking radiation of a black hole. It is also interesting to study the evaporation across a continuous phase transition or crossover \cite{Altman,Balents,our,sk jian,yyz condensation}. Another interesting direction is to study the envolution of entanglement entropy in such systems \cite{para1,para2,para3}.

\textit{Acknowledgements} We thank Xiao-Liang Qi for bringing our attention to the evaporation problem of SYK models and many inspiring discussions. We also want to thank Chao-Ming Jian and Shunyu Yao for helpful discussion. We acknowledge support
from the Walter Burke
Institute for Theoretical Physics at Caltech.

After finishing this work, we became aware that Almheiri, Milekhin, and Swingle have also studied the thermalization of two coupled SYK clusters using Schwinger-Keldysh and exact diagonalization.

\end{document}